\documentclass{emulateapj}
\received{2014 April}
\slugcomment{submitted to {\it the Astrophysical Journal Letters}}

\usepackage{amsmath,amssymb}

\shorttitle{Faint Pop III SNe as the origin of the most iron-poor stars}
\shortauthors{Ishigaki et al.}

\begin{document}


\title{Faint Population III supernovae as the origin of the most iron-poor stars }

\author{Miho N. Ishigaki$^{1}$, Nozomu Tominaga$^{1,2}$, Chiaki Kobayashi$^{1,3}$, and Ken'ichi Nomoto$^{1,4}$}
\affil{$^{1}$Kavli Institute for the Physics and Mathematics of the Universe (WPI), The University of Tokyo, Kashiwa, Chiba 277-8583, Japan; miho.ishigaki@ipmu.jp}
\affil{$^{2}$Department of Physics, Faculty of Science and Engineering, Konan University, 8-9-1 Okamoto, Kobe, Hyogo 658-8501, Japan}
\affil{$^{3}$School of Physics, Astronomy and Mathematics, Centre for Astrophysics Research, University of Hertfordshire, College Lane, Hatfield AL10 9AB, UK}
\affil{$^{4}$Hamamatsu Professor}




\begin{abstract}
The most iron-poor stars in the Milky Way provide important 
observational clues to the astrophysical objects 
that enriched the primordial gas with heavy elements. 
Among them, the recently discovered iron-deficient star 
SMSS J031300.36-670839.3 shows a remarkable chemical composition  
 with non-detection of iron ([Fe/H]$<-7.1$) and large 
enhancement of carbon and magnesium relative to calcium.  
We investigate supernova yields of metal-free (Population III) 
stars to interpret the abundance pattern observed in this star.
We report that the high [C/Ca] and [C/Mg] ratios and upper limits 
of other elemental abundances are well 
reproduced with the yields of core-collapse supernovae (that have 
normal kinetic energies of explosion $E$ of $E_{51}=E/10^{51}$erg$=1$) and 
hypernovae ($E_{51}\geq 10$) of Population III 
25$M_{\odot}$ or 40$M_{\odot}$ stars. 
The best-fit models assume that the explosions
undergo extensive matter mixing and 
fallback, leaving behind a black hole remnant. 
In these models, Ca is produced by 
static/explosive O burning and incomplete Si burning 
in the Population III supernova/hypernova, 
in contrast to 
the suggestion that Ca is originated from the hot-CNO cycle 
during the presupernova evolution. 
Chemical abundances of four carbon-rich 
iron-poor stars with [Fe/H]$<-4.5$, including SMSS J031300.36-670839.3 
are consistently explained by 
 the faint supernova models with the ejected mass of 
$^{56}$Ni less than 10$^{-3}M_{\odot}$.  

\end{abstract}





\section{Introduction}
\label{sec:intro}

Characteristic masses of the first stars 
 (Population III or Pop III stars) and the nature of their 
supernova explosions 
are critically important to determine their role 
in the cosmic reionization and subsequent star formation 
in the early universe \citep[e.g.][]{bromm11}. 
Cosmological simulations have shown 
that the Pop III stars could be very massive $\gtrsim 100M_{\odot}$, 
as the result of cooling of primordial gas via hydrogen molecules \citep
[e.g.][]{bromm04}. 
More recent studies, however, propose the mechanisms in which 
lower mass stars can form through radiation feedback from growing
protostars and/or disk fragmentation 
\citep{hosokawa11,clark11,hirano13,bromm13,susa13}. 

Abundance patterns of the lowest-metallicity stars in 
our Galaxy provide us with a rare opportunity to  
observationally constrain the masses 
of the Pop III stars.   
The chemical abundances in the four iron-poor stars with [Fe/H]$<-4.5$, 
HE 0107-5240 \citep{christlieb02}, HE 1327-2326 \citep{frebel05,aoki06}, 
HE 0557-4840 \citep{norris07}, and SDSS J102915$+$172927 \citep{caffau11} 
(see \citet{hansen14} for a recent discovery of another metal-poor star 
in this metallicity range) do not show signature of 
pair-instability supernovae of very massive 
 ($\gtrsim 140M_{\odot}$) stars as their progenitors  
\citep[][and references therein]{nomoto13}. 
Instead, the observed abundances in these stars 
are better explained by the yields of 
core-collapse supernovae of moderately massive Pop III stars
with several tens of $M_{\odot}$ 
\citep{umeda02,umeda03,limongi03,iwamoto05,tominaga07b,
tominaga09,tominaga14, heger10, kobayashi14}. 

 Another important insight into the nature of 
the Pop III stars is that a large fraction of most iron-poor stars 
are carbon-rich \citep[e.g.][]{hansen14}.  
\citet{iwamoto05} suggests that the large 
enhancement of carbon observed in both HE 0107-5240 
and HE 1327-2326 is explained by their models 
with mixing of supernova ejecta and their 
subsequent fallback on to the central remnant.
These models with different 
extent of the mixing regions simultaneously reproduce 
the observed 
similarity in [C/Fe] and more than a factor of $\sim 10$ 
differences in [O, Mg, Al/Fe] between 
the two stars. 

A metal-poor star SMSS J031300.36-670839.3 (SMSS J0313-6708),  
recently discovered by SkyMapper Southern Sky Survey,  
provides us with a new opportunity to test theoretical 
predictions about the Pop III stars \citep{keller07,keller14}. 
Follow-up spectroscopic observations found that this object 
shows an extremely low upper limit for its iron abundance 
([Fe/H]$<-7.1$); more than $1.0$ dex lower 
than the previous record of the lowest iron-abundance stars. 

In this letter, we extend the study of \citet{iwamoto05} and examine 
whether the abundances of the five stars with [Fe/H]$<-4.5$, 
including the most iron-deficient star SMSS 0313-6708, can be 
consistently explained by the supernova yields of Pop III 
stars which undergo the mixing and fallback.

\section{Models}
\label{sec:model}

We employ the Pop III supernova/hypernova 
yields of progenitors with main-sequence masses of 
$M=25M_{\odot}$ \citep{iwamoto05} and $40M_{\odot}$ 
\citep{tominaga07a} and assume kinetic explosion energies, 
$E$, of $E_{51}=E/10^{51}$erg$=1, 10$ for 
the $25M_{\odot}$ model and 
$E_{51}=1, 30$ for the $40M_{\odot}$ model in the same 
way as in \citet{tominaga07b}. 
The abundance distribution after the explosions of the 
$25M_{\odot}$ models with $E_{51}=1$ and $10$
as a function of the  
enclosed mass ($M_{r}$) are illustrated in Figure \ref{fig:model}.

For the parameters that describe the extent 
of the mixing-fallback, we follow the prescription of 
\citet{umeda02} and \citet{tominaga07b} as briefly summarized below. 
We assume that the supernova ejecta within a $M_{r}$ range between   
an initial mass-cut $M_{\rm cut}(\rm ini)$ and 
$M_{\rm mix}(\rm out)$ are mixed and a fraction $f$ of the mixed material
fall back onto the central remnant, presumably forming a black hole. 
This approach has been compared with a 
hydrodynamical calculation of jet-induced explosions \citep{tominaga09}
and it is demonstrated that the prescription of the mixing-fallback
applied to a one-dimensional calculation mimics 
an angle-averaged yield of the aspherical supernovae. 

We assume that the $M_{\rm cut}(\rm ini)$ is approximately located at the 
outer edge of the Fe-core \citep[$M_{r}=1.79 M_{\odot}$ and $2.24 M_{\odot}$
for $M=25 M_{\odot}$ and $40 M_{\odot}$, respectively: ][]{tominaga07b}. 
The value of $M_{\rm cut}({\rm ini})$ is 
varied within $\pm 0.3 M_{\odot}$ with steps of 0.1 $M_{\odot}$ 
to better fit the observed 
abundance patterns. Then we calculate the grids of 
supernova yields for the range of $\log f=-7.0$-$0.0$
with steps of $\Delta\log f=0.1$ and for
$M_{\rm mix}({\rm out})=1.5$-$9.0 M_{\odot}$ ($M=25M_{\odot}$) 
and 2.0-16.0 $M_{\odot}$ ($M=40M_{\odot}$) with steps 
of $\Delta M_{\rm mix}({\rm out})=0.1M_{\odot}$.   
These ranges have been chosen so that the $M_{\rm mix}({\rm out})$ 
is searched approximately in the range between the location of the $M_{\rm cut}({\rm ini})$ 
and the outer boundary of the CO core (see Figure \ref{fig:model}). Using this
grid of yields, parameter sets ($M_{\rm cut}({\rm ini})$, $M_{\rm mix}({\rm out}), f$) that reproduce 
the observed [C/Ca] and [C/Mg] ratios in SMSS 0313-6708 are searched.
In addition to SMSS 0313-6708, we adopt the same parameter-search 
method for the four other iron-poor stars with [Fe/H]$<-4.5$ 
(see \citet{tominaga14} for models with other $E$). 

\begin{figure*}
\begin{center}
\includegraphics[width=13.0cm]{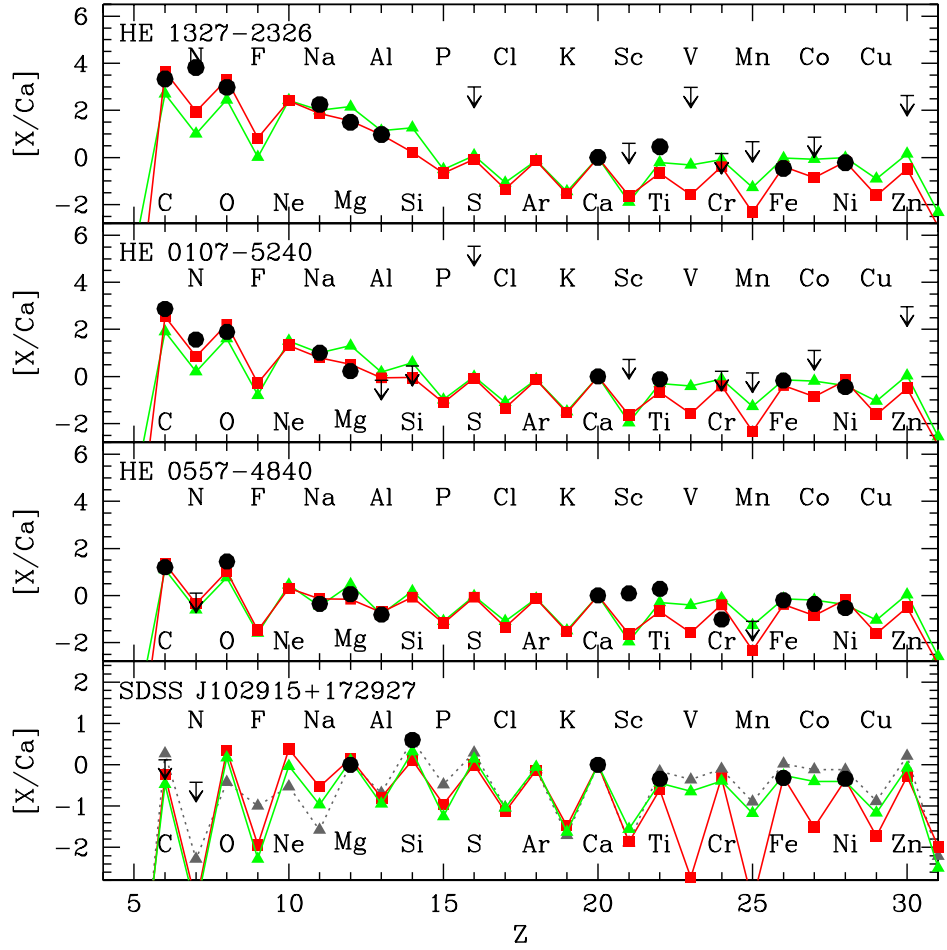}
\caption{ Elemental abundance patterns (relative to Ca derived from Ca I lines)
 of the best-fit models for the four other iron-poor stars with [Fe/H]$<-4.5$. 
The squares (red) and triangles 
(green) show supernova and hypernova models, respectively, with $M=25M_{\odot}$ 
($E_{51}=1$ or $10$) for the three carbon-enhanced stars 
(top-three panels) and $M=40M_{\odot}$ ($E_{51}=1$ or $30$) for the 
star without carbon enhancement (the bottom panel). The gray triangles 
with dotted line in the bottom panel indicate the hypernova model with 
an alternative set of parameters ($M_{\rm mix}(\rm out)=13.9$, $\log f=-1.9$). 
Observational data sources are; 
HE 0107-5240: \citet{collet06} with 3D-LTE, HE 1327-2326: \citet{frebel08} 
with 3D-LTE and \citet{bonifacio12} for S with 3D-NLTE, HE 0557-4840: \citet{norris07} and \citet{norris12} with 3D-LTE, SDSS J102915$+$172927: 
\citet{caffau12} with 3D-NLTE. 
\label{fig:hmpump} }
\end{center}
\end{figure*}

We adopt the observational data analyzed with 3D and/or NLTE corrections 
(see the captions of Figure \ref{fig:abupattern} and \ref{fig:hmpump} 
for details) 
 and normalized 
with the solar abundances of \citet{asplund09}. 
For Ca abundances, we use abundances estimated from Ca I lines 
for the four other iron-poor stars. 
We should note, however, that there are uncertainties in the Ca abundances
for the 3D and NLTE effects \citep[e.g.][]{korn09,lind13}, which affects the 
normalization of Figure \ref{fig:hmpump}. 
We also consider Ca II abundances when we draw our conclusions. 

 \section{Result}
\label{sec:result}

Table \ref{tab:bestmodels} summarizes the model parameters 
($M_{\rm cut}({\rm ini})$, $M_{\rm mix}({\rm out})$, $f$) that best 
reproduce the observed abundance patterns of the five iron-poor stars. 
The resultant 
masses of the central remnant ($M_{\rm rem}$) and ejected masses 
of $^{56}$Ni, which
finally decays to $^{56}$Fe, are indicated in the last two columns.  
Since only C, Mg, and Ca have been measured for SMSS 0313-6708, 
all four models ($(M, E_{51})=(25, 1)$, $(25, 10)$, $(40, 1)$, and 
$(40, 30)$) can equally well fit the observed abundance ratios as 
indicated in Table \ref{tab:bestmodels}.  
For the other four stars, the parameters have been constrained using 
a larger number of elements and thus only the best-fit models 
are shown. 

\subsection{The best-fit models for SMSS 0313-6708 \label{sec:bestfit}}

\begin{figure*}
\begin{center}
\includegraphics[width=13.0cm]{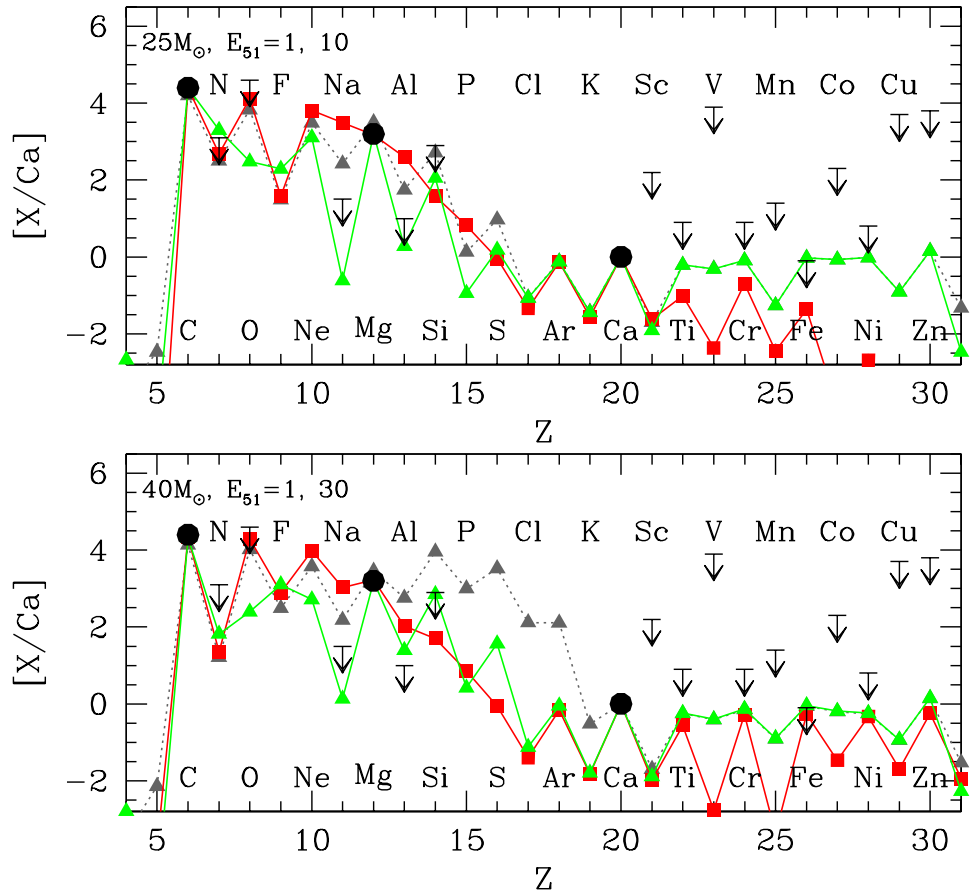}
\caption{ Elemental abundance patterns (relative to Ca)
of the best-fit models for SMSS 0313-6708
(observational data from \citet{keller14} with 3D-NLTE: 
filled circles and arrows for the upper limits) for 
 $M=25M_{\odot}$ (top) and 40$M_{\odot}$ (bottom). 
The solid lines with squares (red) show the supernova model with
$E_{51}=1$. The solid lines with triangles (green) 
show the hypernova model with 
$E_{51}=10$ (for $M=25M_{\odot}$) and $E_{51}=30$ (40$M_{\odot}$). The 
dotted lines with triangles (gray) indicate another hypernova model with 
different parameters from the best-fit model 
($M_{\rm mix}({\rm out})=5.7M_{\odot}$ and $\log f=-5$).
\label{fig:abupattern}}
\end{center}
\end{figure*}

Figure \ref{fig:abupattern} shows 
the abundance patterns relative to Ca of 
the best-fit models 
for $M=25M_{\odot}$ 
(top) and 40$M_{\odot}$ (bottom). 
In each panel, the models of the supernova ($E_{51}=1$: 
solid lines with squares) and hypernovae ($E_{51}=10$ and 30 for 
$M=25M_{\odot}$ and 40$M_{\odot}$: solid lines with triangles) are shown 
with the observed abundances in 
SMSS 0313-6708 (filled circles and arrows for the upper limits).
In the following we describe comparison 
of each model with the observed abundance pattern.

{\it $M=25M_{\odot}$, $E_{51}=1$, supernova model: }
 The model yield that fits the observed 
[C/Ca] and [Mg/Ca] ratios are consistent with 
the upper limits of other elements except for 
Na and Al. Because the predicted Na and Al abundances vary by more than 
a few dex depending on overshooting \citep{iwamoto05} 
and stellar rotation \citep{meynet10}, and on the reaction rate of 
$^{12}$C$(\alpha, \gamma)$$^{16}$O \citep[][and references therein]{chieffi02} 
in the presupernova models, 
we restrict our discussion for Na and Al to 
their relative yields between the supernova and hypernova models.  

\begin{figure*}[h]
\begin{center}
\includegraphics[width=13.0cm,angle=270]{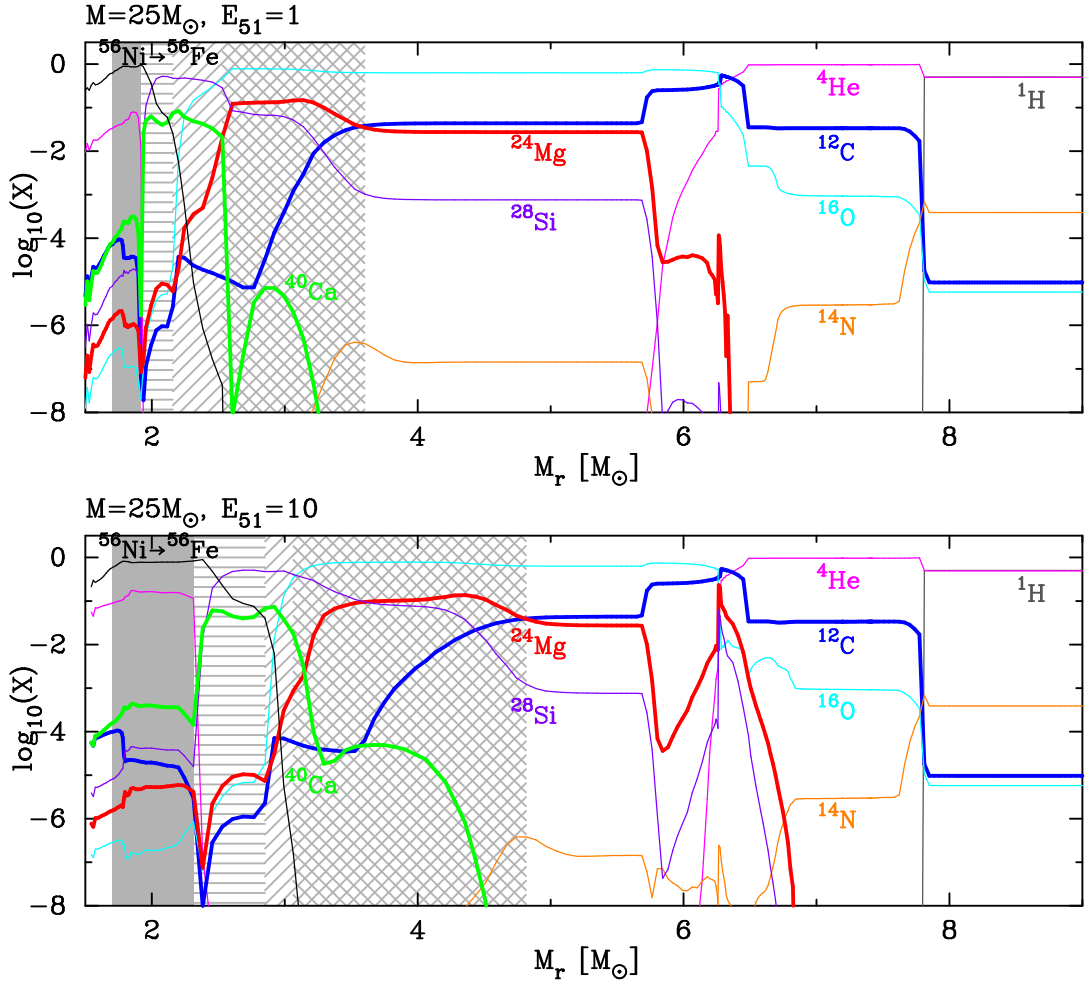}
\caption{(Top) The distribution of elemental abundances as a 
function of the enclosed 
 mass ($M_{r}$) for a model with $M=25M_{\odot}$ and $E_{51}=1$.
Regions where explosive burning takes places are shown with 
filled (complete Si burning), bordered (incomplete Si burning), hatched (O burning), 
and cross-hatched (C burning) areas. (Bottom) Same as the top panel 
but for a model with $M=25M_{\odot}$ and $E_{51}=10$. \label{fig:model} }
\end{center}
\end{figure*}

In our model, the observed large [Mg/Ca] ratio 
results from the large mixing region 
($M_{\rm mix}({\rm out})=5.6M_{\odot}$) and 
the small ejected fraction, $f$;  
as seen in Figure \ref{fig:model}, the material in the 
layer containing Ca is mostly falls back while 
the material in the outer layer containing Mg is partly ejected.   
The abundances of iron-peak elements in the model depend on 
 the adopted $M_{\rm cut}({\rm ini})$.
For $M_{\rm cut}({\rm ini})=2.0 M_{\odot}$, 
the ejected mass of $^{56}$Ni is less than 
 $10^{-5}M_{\odot}$ for these models,  
which is extremely 
small compared to those estimated for nearby 
supernovae \citep[$^{56}$Ni$\sim 0.1M_{\odot}$: ][]{blinnikov00,nomoto13}.

{\it $M=25M_{\odot}$, $E_{51}=10$, hypernova model: }
The higher explosion energy induces explosive 
burning at the bottom of the He layer, which leads to 
an extra production of Mg at $M_{r}\sim 6M_{\odot}$ as 
can be seen in the bottom panel of Figure \ref{fig:model}. 
Consequently, a large value of $M_{\rm mix}({\rm out})$ 
($6.4M_{\odot}$), which results in a large fallback of Mg 
synthesized at $M_r\sim6M_\odot$, 
best explain the observed [C/Mg] ratio.  
The [O/Ca] ratio is smaller than that of the supernova 
model as a result of the associated fallback of oxygen for the given [Mg/Ca] 
ratio. The [O/Ca] ratio, however, can be as large as $\sim +4$ 
depending on $M_{\rm mix}({\rm out})$ and $f$ as shown by 
the dotted gray line in Figure \ref{fig:abupattern}.

The [Na/Ca] and [Al/Ca] ratios of the hypernova yields 
are lower than those predicted by the supernova model. 
This is due to the more extended explosive 
O and C burning regions in the more energetic 
explosion [cross-hatched regions in 
Figure \ref{fig:model}, in which more Na and Al are consumed to synthesize 
heavier nuclei \citep{nakamura01}]. The larger 
$M_{\rm mix}({\rm out})$ with small $f$  
also leads to the smaller amount of ejected Na and Al than the 
supernova model.

The abundance differences between the odd and even atomic number 
iron-peak elements are smaller for the hypernova model. 
This results from enhanced production of 
odd-Z iron-peak elements due to the higher entropy
and larger neutron excess in the Si burning region in the 
more energetic explosion \citep[e.g.][]{nakamura01}. 

{\it $M=40M_{\odot}$, $E_{51}=1$, supernova model: }
Compared to the supernova model with $M=25M_{\odot}$, 
the outer boundary of the Mg-rich layer extends to a larger 
mass at $M_r\sim 13M_{\odot}$. 
Consequently, the models with 
$M_{\rm mix}({\rm out})\sim 12-14M_{\odot}$ best-fit the observed [C/Mg] ratio.

{\it $M=40M_{\odot}$, $E_{51}=30$, hypernova model: }
The hypernova model for the
$40M_{\odot}$ progenitor is characterized by 
larger [Si/Ca] and [S/Ca] ratios because of more 
extended explosive O and C-burning region, in which elements such 
as O and C are consumed to synthesize Si and S. Aluminum 
is also synthesized in the 
explosive C burning layer but destroyed in the explosive O burning layer 
to synthesize Si. 
In comparison to the 
supernova model, predicted abundances of V, Mn, Co and 
Cu relative to Ca are larger, similar to
the $25M_{\odot}$ hypernova model.  

\begin{table*}
\begin{center}
\caption{Summary of the observed abundances and the best-fit models \label{tab:bestmodels}}
\begin{tabular}{lcccccccccc}
\hline
Name  & [Fe/H]\tablenotemark{a}&[C/Ca]\tablenotemark{a} &[C/Mg]\tablenotemark{a} &  $M$ & $E_{51}$&  $M_{\rm cut}$(ini) & $M_{\rm mix}$(out)  & $\log f$& $M_{\rm rem}$ &  $M(^{56}{\rm Ni})$ \\
      & (dex) & (dex) & (dex) & ($M_{\odot}$)&($10^{51}$erg) &  ($M_{\odot}$) & ($M_{\odot}$)  &    & ($M_{\odot}$) &  ($M_{\odot}$)  \\
\hline
 SMSSJ0313-6708       & $<-7.1$ & 4.4 & 1.2 & 25 & 1   &  2.0  &  5.6   & $-5.1$  &  5.6    & 2.1$\times 10^{-7}$    \\
                       &         &  & & 25 & 10  &  1.7  &  6.4   & $-5.8$  &  6.4    & 1.1$\times 10^{-6}$  \\
                       &         &  & & 40 & 1   &  2.0  & 12.7   & $-5.4$  & 12.7    & 3.4$\times 10^{-6}$  \\
                       &         & & & 40 & 30  &  2.5  & 14.3   & $-6.0$  & 14.3    & 1.9$\times 10^{-6}$  \\
 HE0107-5240          & $-5.7$  &2.9&2.6 & 25 & 1   &  1.7  &  5.7   & $-3.3$  &  5.7    & 1.4$\times 10^{-4}$    \\
 HE1327-2326          & $-6.0$  &3.3 &1.8 & 25 & 1   &  1.7  &  5.7   & $-4.4$  &  5.7    & 1.1$\times 10^{-5}$    \\
 HE0557-4840          & $-4.9$  & 1.2 & 1.1& 25 & 1   &  1.7  &  5.7   & $-2.1$  &  5.7    & 2.2$\times 10^{-3}$    \\
 SDSSJ102915$+$172927 & $-4.9$  & $<0.1$& $<0.1$ & 40 & 30  &  2.0  & 5.5   & $-0.9$  &  6.0   & 2.8$\times 10^{-1}$    \\
\hline
\end{tabular}
\tablenotetext{1}{See the captions of Figure \ref{fig:abupattern} 
and \ref{fig:hmpump} for the references of the observational data.}
\end{center}
\end{table*}

\subsection{Four other iron-poor stars with [Fe/H]$<-4.5$ }

Figure \ref{fig:hmpump} shows the abundance patterns of 
the best-fit models for four other iron-poor stars with [Fe/H]$<-4.5$. 
We note that the N abundances are not included from the fitting because of 
their uncertainty in the progenitor models \citep{iwamoto05}.
For the three carbon-enhanced stars 
(HE 1327-2326, HE 0107-5240, and HE 0557-4840) 
shown in the top-three panels,  
the supernova models of $M=25 M_{\odot}$ with $E_{51}=1$
adopting the parameters $M_{\rm mix}({\rm out})\sim 5.7 M_{\odot}$ and 
$f\sim 10^{-4}-10^{-2}$ 
best explain the observed 
abundances (squares). 
The hypernova models with $E_{51}=10$ (triangles)
tend to yield higher [Mg/Ca] ratios than the observed values, 
which is due to the explosive synthesis of Mg at the 
bottom of the He layer. 

 The abundance pattern of SDSS J102915$+$172927 with no 
carbon enhancement is in better agreement with the hypernova yields of 
$M=40M_{\odot}$ and $E_{51}=30$ (triangles). 
This is due to the relatively high [Si/Ca] ratio 
observed in this star, which is better explained by the 
larger Si/Ca ratio in the higher explosion energy model.
Such an energy dependence can be understood from the comparison of the
abundance distributions between the top and bottom panels of Figure
\ref{fig:model}. In the hypernova model (bottom), 
the region where the postshock
temperature becomes high enough to produce a significant amount of Si
extends to larger $M_r$ than in the supernova model (top). 

\subsection{Comparison of the five most iron-poor stars}

\begin{figure*}
\begin{center}
\includegraphics[width=13.0cm]{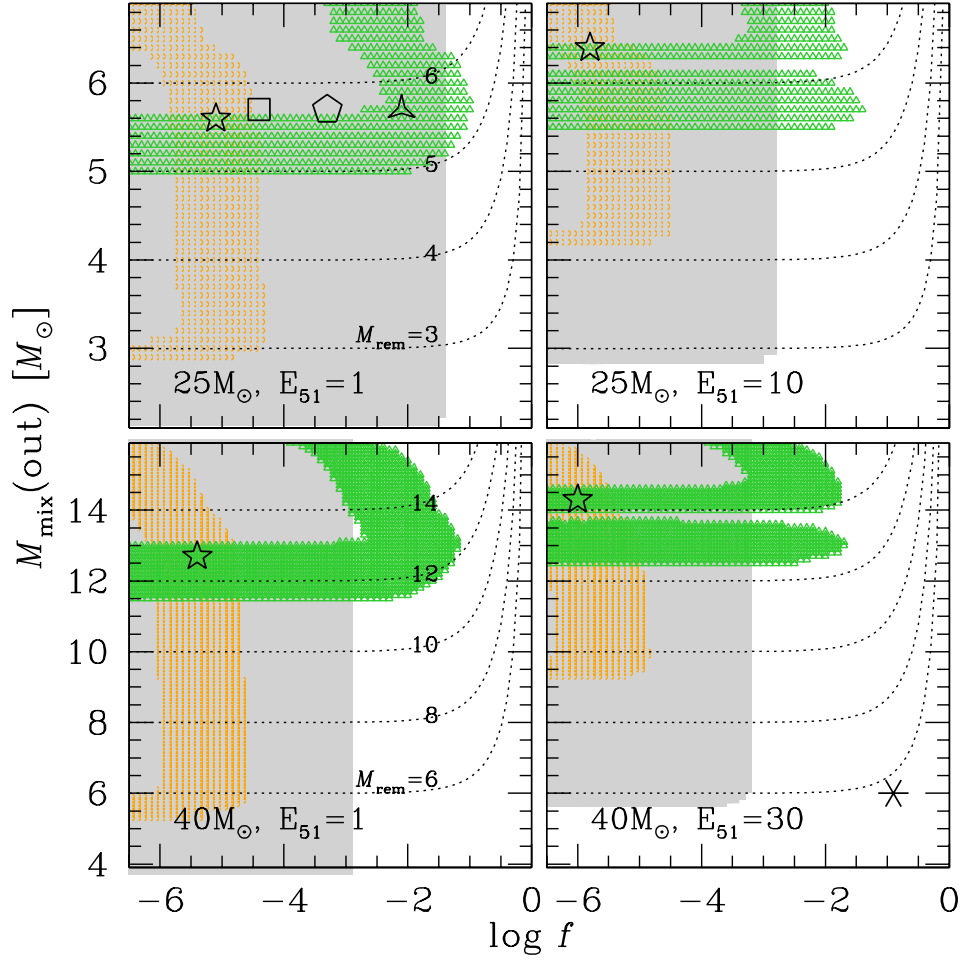}
\end{center}
\caption{The regions in the $M_{\rm mix}({\rm out})-\log f$ parameter 
space for the observed [C/Ca] (orange) and [C/Mg] (green) ratios 
within $\pm 3\sigma$ errors. The best-fit parameters
for SMSS 0313-6708 in each model are shown with stars. 
The best-fit models for HE 1327-2326 (square), HE 0107-5240 (pentagon), 
HE 0557-4840 (three-pointed star), and SDSS J102915$+$172927 (asterisk) are 
also shown. The dotted lines indicate the remnant mass in the 
corresponding parameter set. 
The region which represents a relatively faint 
supernova with $M (^{56}{\rm Ni})< 10^{-3} M_{\odot}$ is shown by gray 
area \label{fig:c_ca_mg}.}
\end{figure*}

Figure \ref{fig:c_ca_mg} summarizes the best-fit 
parameters ($M_{\rm mix}({\rm out}), \log f$) obtained for the 
five iron-poor stars 
(star: SMSS 0313-6708, square: HE 1327-2326, pentagon: HE 0107-5240, 
three-pointed star: HE 0557-4840, and asterisk: SDSS J102915$+$172927). 
The results for the model 
$(M, E_{51})=(25, 1), (25, 10), (40, 1),$ and $(40, 30)$ are shown from 
the top-left to bottom-right panels. 
The regions where the [C/Ca] and [C/Mg]
 ratios measured within $\pm 3\sigma$ in SMSS 0313-6708 are realized  
 are shown by orange and green, respectively.

In the $M_{\rm mix}({\rm out})-\log f$ parameter space, the [C/Ca] ratio 
is at maximum when $M_{\rm mix}({\rm out})$ is located at 
the layer where carbon burning takes place ($M_r\sim 3-4M_{\odot}$ in the
$M=25M_{\odot}$ and $E_{51}=1$ model; top panel of Figure \ref{fig:model}).  
At a given $M_{\rm mix}({\rm out})$, the [C/Ca] 
is larger for smaller $f$ (larger fallback), because Ca 
synthesized in the inner layer falls back more efficiently for 
smaller $f$ than C synthesized in the outer layer.
As a result, the high [C/Ca] ratio (orange in Figure \ref{fig:c_ca_mg}) 
in SMSS 0313-6708 is reproduced 
with an extensive fallback ($\log f<-4$). 

The [C/Mg] ratio is at maximum when the $M_{\rm mix}({\rm out})$ 
is located at the outer edge of the Mg-rich ejecta. 
The ejection of a large amount of C and the fallback of 
most of the Mg-rich layer lead to the large [C/Mg] ratios.
The observed [C/Mg] ratio in 
SMSS 0313-6708 is close to the maximum for the observed [C/Ca].  
 Consequently, mixing up to $M_r \sim 6M_{\odot}$ 
(for $M=25M_{\odot}$) and $\sim 13M_{\odot}$ (for 
$M=40M_{\odot}$) and small $f$ are required 
(green in Figure \ref{fig:c_ca_mg}). 
We note that the gap 
in the [C/Mg] constraint, which is only seen in the hypernova models, 
stems from the explosive Mg production in the energetic explosion 
(Section \ref{sec:bestfit}). The resultant [C/Mg] ratio at the 
gap is lower than the observed [C/Mg] ratio. 

For the three other iron-poor stars showing 
carbon enhancement, the best-fit value of $M_{\rm mix}({\rm out})$ 
is as large as that required for
SMSS 0313-6707 with $\log f\leq -2$. 
The obtained parameters
$M_{\rm mix}({\rm out})$ and $f$
remain similar if we take the average of Ca I and Ca II abundances 
instead of Ca I abundances, which gives 0.2 dex systematically 
lower [C/Ca] than in Table 1.
They also remain similar  
if only C, Mg, and Ca abundances are taken into 
account in the fitting. This is because these elements 
are synthesized in the different layers and dominant in each layer 
(the thick lines in Figure \ref{fig:model}) and 
thus their abundance ratios are
mostly determined by the mixing-fallback.  

The abundance pattern of SDSS J102915$+$172927, 
which does not show carbon enhancement, is best reproduced by the 40$M_{\odot}$ 
model with $M_{\rm mix}({\rm out})= 6.0 M_{\odot}$ and $\log f=-0.9$ (asterisk in
Figure \ref{fig:c_ca_mg}). If we use the average of Ca I and Ca II abundances, 
a model with $M_{\rm mix}({\rm out})= 13.9 M_{\odot}$ and $\log f=-1.9$ 
best reproduce the observed abundances 
(gray triangles with the dotted line in the 
bottom panel of Figure \ref{fig:hmpump}).

 The differences in masses, energies and the state of mixing-fallback 
in the fitted models may explain populations 
with and without carbon enhancement in the Galactic halo stars 
\citep[e.g.][]{norris13}. Analyses 
based on a larger number of sample are required to examine
whether or not the properties of Pop III supernovae/hypernovae can reproduce 
the abundance patterns and the relative fraction of the C-enhanced
and the C-normal metal-poor populations 
in the Galactic halo. 

The gray-filled region in Figure \ref{fig:c_ca_mg} represents 
the range of the parameters that 
give the ejected Fe mass less than 
$10^{-3} M_{\odot}$, which presumably corresponds to faint Pop III supernovae. 
All three stars showing the C enhancement are located 
in the faint supernova region.

\section{Discussion}
\label{sec:discussion}

As shown in the previous section, the 
abundance measurements (C, Mg, and Ca) in 
SMSS 0313-6707 are reproduced with 
the Pop III supernova/hypernova yields 
($(M, E_{51})=(25, 1), (25, 10), (40, 1)$ and $(40, 30)$),
 with the model
parameters corresponding to faint supernova/hypernova with
extensive mixing and prominent fallback. 
In order to discriminate 
models with different masses and energies, additional 
abundance measurements for oxygen as well as 
iron-peak elements including V, Mn, Co, and Cu
are particularly useful.

The ejected mass of $^{40}$Ca is only  
$\sim 10^{-7}-10^{-8} M_{\odot}$ in the faint supernovae/hypernovae 
models. To be compatible with 
the observed calcium abundance ([Ca/H]$=-7.0$), 
 the supernova ejecta should be diluted 
with $\sim 10^{3}- 10^{4}M_{\odot}$ of the primordial gas. 
In the case of the supernova models with $E_{51}=1$, this is 
consistent with the suggested relation between the supernova energy 
versus swept-up gas mass with primordial composition \citep{thornton98}, 
 as adopted in \citet{tominaga07b}, 
for the assumed number density of hydrogen $1<n_{H}<100$ cm$^{-3}$. 
On the other hand, this relation predicts that the hypernova 
sweeps up much larger amount of hydrogen 
 ($\gtrsim 10^{5}M_{\odot}$ for $n_{H}<10^4$ cm$^{-3}$) 
than the above values.
A recent cosmological simulations of the transport of the metals 
synthesized in a Pop III supernova, however, 
suggests a wide range of 
metal abundances in the interstellar gas clouds 
after an explosion of a Pop III star \citep{ritter12},
 and thus the hypernova 
could work as the source of the chemical 
enrichment for the formation of stars with [Ca/H]$\lesssim -7$.

In our model, Ca is produced by hydrostatic/explosive O burning 
and incomplete Si burning in the Pop III supernova or 
hypernova with masses $25$ or $40 M_{\odot}$.
This is different from the 60$M_{\odot}$ model adopted in \citet{keller14}, 
where Ca is originated from the hot CNO cycle during the main-sequence phase.
Synthesis of $\sim 10^{-7}M_{\odot}$ Ca in the hot CNO cycle 
is also seen in the $100 M_{\odot}$ models of \citet{umeda05}.
On the other hand, Ca produced in this mechanism is not seen  
in the 25 and 40$M_{\odot}$ progenitors analyzed in this paper. 
The mass fraction of Ca near the bottom of the hydrogen layer in 
these progenitors is 
$\log X_{\rm Ca}<-10$.

In order to clarify which of these nucleosynthesis sites are 
responsible for the observed Ca, we note a different prediction 
between the two scenarios. 
Our models suggest that a certain amount of Fe distributed in the 
inner region as well as explosively synthesized 
Ca are ejected as a result of the assumed mixing at $M_{r}=2-6M_{\odot}$. 
This results in the [Fe/Ca] ratio of $\sim -2$ - $0$, 
depending on the $M_{\rm cut}({\rm ini})$. Consequently,  
our models of the faint supernova/hypernova
predict the metallicity distribution 
function (MDF) to be continuous even below [Fe/H]$<-6$.  
On the other hand, the model adopted in \citet{keller14}, in which 
Ca is produced in the hot-CNO cycle, 
predicts [Fe/Ca]$\lesssim-3$, which is not observed in 
other extremely iron-poor stars. Because of the distinct 
Ca production sites, the MDF could be 
discontinuous in the most metal-poor region.  
 Future photometric and spectroscopic surveys to 
discover lowest metallicity stars and their MDF provide clues to discriminate 
these mechanisms. 

The models adopted 
in this work suggest that the faint Pop III supernovae could be the origin of 
the observed abundance patterns and the variation among the 
most iron-poor carbon-rich stars. 
To understand physics of faint Pop III supernovae, 
multi-dimensional simulations are necessary.  
A large-scale mixing as suggested for the carbon-enhanced stars are
 not predicted in the models with Rayleigh-Taylor mixing alone 
\citep{joggerst09}.
Instead, a more likely origin of such large-scale mixing-fallback would be 
the jet-induced supernova/hypernova, where the inner material can be ejected 
along the jet-axis while a large fraction of the material along 
the equatorial 
plain falls back \citep{tominaga09}.

\acknowledgements

This work has been supported by the Grant-in-Aid for JSPS fellow
(M.N.I.) and for Scientific Research of the JSPS (23224004, 23540262,
26400222), and by WPI Initiative, MEXT, Japan. 
 The authors thank
S. Keller, M. Asplund, and K. Lind for fruitful discussion.

\end{document}